\documentclass[runningheads]{llncs}

\usepackage[T1]{fontenc}
\usepackage{comment}
\usepackage{graphicx}
\usepackage{xcolor}

\begin{document}
 
\title{JuPyLive: Seamless Migration of Jupyter Notebook Resources from Laptop to HPC}

\author{Sima Attar-Khorasani\inst{1,2}\orcidID{0000-0002-7396-1983} \and
Matthias Lieber\inst{1,2}\orcidID{0000-0003-3137-0648} \and
Siavash Ghiasvand\inst{1,3}\orcidID{0000-0001-6627-0159}}
\authorrunning{Sima Attar-Khorasani, Matthias Lieber, Siavash Ghiasvand}

\institute{Center for Interdisciplinary Digital Sciences (CIDS), TUD Dresden University of Technology, Dresden, Germany \and
Center for Information Services and High Performance Computing (ZIH), TUD Dresden University of Technology, Dresden, Germany \and
Center for Scalable Data Analytics and Artificial Intelligence (ScaDS.AI), Dresden/Leipzig, Germany}

\maketitle              

\begin{abstract}

This work introduces JuPyLive, a migration mechanism that enables seamless transition of Jupyter notebooks between local resources of user's workstation and remote resources of high-performance computing~(HPC) environments, while preserving the user experience.
JuPyLive eliminates the underlying complexities of migration process, enabling users to freely choose among available local and remote resources, directly within the familiar Jupyter notebook environment via a single click.
JuPyLive leverages ElasticNotebook to manage in-memory state migration, it automates resource allocation on HPC cluster and orchestrates required remote communication channels between the source and destination to enable a bidirectional migration.
Furthermore, HPC status monitor of JuPyLive provides a live overview of available remote resources, allowing users to make informed decisions on choosing the relevant resources before initiating a migration process.
The proposed fully automatic mechanism requires no code changes or configurations by the end user, nor does it demand users to learn a new syntax, instead the migration process can be intuitively initiated and monitored using visual elements from within the Jupyter notebook.
By bridging the gap between local workspace and remote resources, JuPyLive offers a seamless experience for scaling local resource-intensive workflows with minimal user intervention, thus further democratizing the usage of HPC clusters among the interdisciplinary researchers.

\keywords{Heterogeneous Live Migration  \and Jupyter Notebook \and High performance Computing (HPC)}
\end{abstract}

\section{Introduction}
\label{sec:intro}
New algorithms are often sophisticated combination of several multi-domain methodologies that employ multiple data processing layers, thus increases the need for higher computational resources.
The new advancements in the field of machine learning and artificial intelligence, and the current shift towards data-driven analysis further increase this demand.
Modern computational models also require more resources to provide higher accuracy and detail.
Many modern machine learning frameworks hide complex computations behind simple interfaces, but the underlying processes remain resource intensive.
These create a paradox in which interdisciplinary researchers can easily implement sophisticated computer programs without computer science background, however they often can run only on increasingly powerful hardware than their predecessors.
Therefore, lowering the access barrier to high performance computing (HPC) resources is an essential step in enabling scientists from various domains to benefit from the potentials of recent advancements, particularly in the field of machine learning.

However, HPC resources are limited and in comparison to local less-powerful resources, significantly more expensive.
Therefore, the access to HPC resources needs to be regulated and carefully considered in order to avoid unnecessary overheads. 

In response to this limitation, a common practice in implementation cycle of computer programs (hereafter: code) is utilizing less powerful local resources for implementing and debugging the early prototypes, and later executing the more mature and tested versions of the code on high performance resources.
However, often certain sections of even early prototypes, such as sections that rely on data-driven algorithms or machine learning frameworks, require more resources than those locally available (e.g., high-end GPUs).
Which in turn requires continuous switching between different computational resources, even for development of early prototypes.

Although this switch between local and HPC resources is a common practice among experienced users, this is not convenient and for many users this migration of code and development environment, which usually requires manual configurations and adjustments, imposes significant overhead and can become a preventive barrier.

Furthermore, most research-oriented HPC clusters are using Slurm as the main job scheduler~\cite{HyperionResearch2023} beside other accounting tools to regulate the access to HPC resources.
The batch nature of job schedulers on HPC clusters, in contrast to the common interactive paradigm of programming, imposes further challenge for a barrier-free access to HPC resources.

To addresses these challenges, this works proposes JuPyLive, a transparent migration mechanism which enables users to seamlessly switch between local and HPC resources on demand; emulating an elastic resource pool that adapts itself based on the users resource demands.
The source code and a demonstration video is available on the code repository~\footnote{Source code: \url{https://gitlab.hrz.tu-chemnitz.de/siat527e{-}{-}tu-dresden.de/jupylive},\\Demo video: \url{https://youtu.be/JnmCyRrKOEE}}.

Considering that the focus of this work is on development environments, the popularity of Jupyter notebook among interdisciplinary researchers, the layers of abstraction and flexibility that it offers, and the cell-based execution context, JupyterHub has been chosen as the underlying development environments.
Although the same mechanism can be implemented in several other development environments.

JuPyLive instances can be initiated on the user's computer or any favorable platform, and whenever user requires different computational resources, a single click performs all necessary adaptations and informs the user once the migration is finished and requested resources are available.
The migration process can be performed as many time as necessary based on the users demand.
It is worth emphasizing that JuPyLive enables on-demand access to all HPC resources such as GPUs within the same Jupyter notebook, even when such resources are not present in the local environment.

The rest of this document is structured as follows: section~\ref{sec:relatedWork} discusses some related work, then in section~\ref{sec:implementation} we will go through the implementation details, section~\ref{sec:assumptions} shows the limitation as well we the assumptions that we set, and finally in section~\ref{sec:conclusion} we'll see Conclusion and Future Direction.

\section{Related Work}
\label{sec:relatedWork}
Several research studies and tools have been developed regarding live migration of processes and containers, as well as virtual machines (VMs).
At the operating system (OS) level, tools such as CRIU(Checkpoint/Restore In Userspace), DMTCP(Distributed MultiThreaded CheckPointing), and FIT(Fault Tolerance Interface), have been introduced.
These approaches have limitations, such as not supporting GPU-accelerated workloads, being very platform dependent, or being unable to handle migration between heterogeneous environments~\cite{Attar2025}.
Furthermore, existing approaches require manual configurations by users and certain expertise beyond the main focus of interdisciplinary researchers.
Which in turn makes their usage impractical for most users.
Among major development environments, Jupyter Hub is highly popular, and thanks to its client-server architecture, it provides an accessible and flexible all-in-on environment, and is accessible to users directly from a web interface, effectively shifting all tedious manual configurations out of user perspective, making Jupyter hub an intuitive platform for most interdisciplinary researchers.
Therefore, some research groups have focused on the live migration of Jupyter notebooks.
At the process level, similar to the container level, researchers in \cite{Juric2021} created ELSA for the live migration of Jupyter notebooks based on Podman and leveraging CRIU.
ELSA migrates Jupyter notebooks running on a CPU to a larger or smaller machine or just to halt them temporarily.
The embedded button in the Jupyter user interface(UI), allows users to seamlessly migrate their workspaces back and forth between nodes with less or more resources via shared NFS storage.
However, since CRIU does not support GPU workloads, this method cannot be used for most current approaches which highly depend on GPU resources.
Research has also been conducted regarding the migration of Jupyter notebooks at the application level.
Selective cell migration in~\cite{Cunha2021} uses AST-based (Abstract Syntax Tree) pruning to optimize hardware-intensive cell executions on hybrid clouds.
However, it assumes complete similarity in libraries and versions between the source and destination, which highly restricts its flexibility.
ElasticNotebook~\cite{elasticNotebook} on the other hand facilitated the migration of Jupyter notebooks by leveraging a cost-based optimization model that chooses between transferring the state and cell recomputation.
A transparent data layer between the UI and the ipykernel intercepts cell execution and tracks the resulting modifications. Our work utilizes the underlying mechanisms introduced by ElasticNotebook to handle the migration.
Furthermore, to the best of our knowledge there are no existing work that integrates batch schedulers such as Slurm in the migration process.

\section{System Overview and Implementation}
\label{sec:implementation}
\begin{figure}
    \centering
    \includegraphics[width=0.5\textwidth]{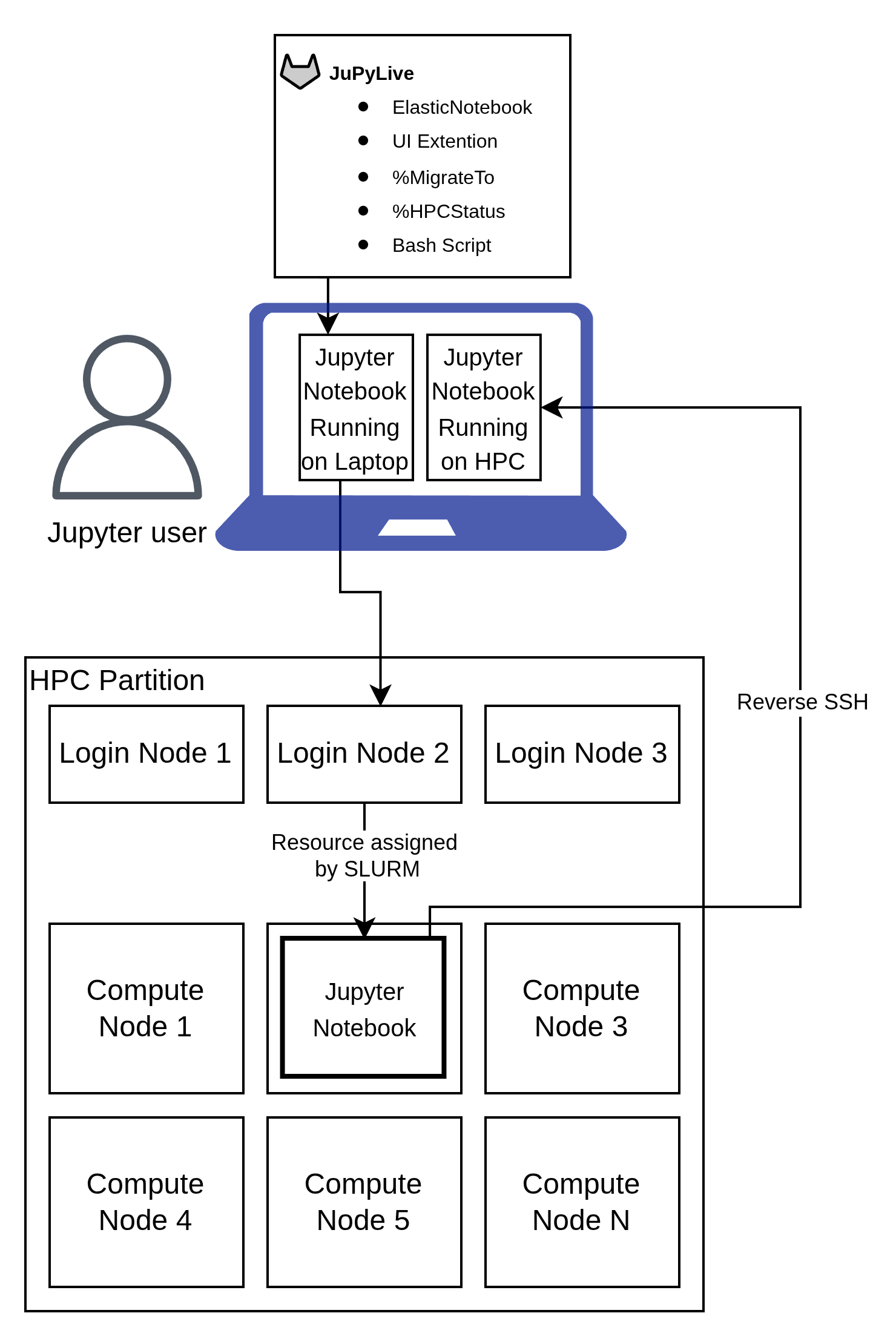}
    \caption{Schematic overview of JuPyLive's workflow}
    \label{fig:workflow}
\end{figure}

JuPyLive is designed to provide a seamless migration between a local workstation and HPC clusters, such that the process of code development can be initiated on the local resources and in case of demand for more resources the development can be seamlessly continued on more powerful resources.
Once the local resources are sufficient, the reverse migration can be performed and development can be seamlessly continued on local resources.
Figure~\ref{fig:workflow} illustrates a schematic overview of the JuPyLive's workflow.

Current implementation leverages the ElasticNotebook framework~\cite{elasticNotebook}, which provides a checkpoint/restore mechanism for Jupyter sessions.
A fully automatic installer prepares the environment and executes the JuPyLive environment.
JuPyLive automates and transparently handles seamless bidirectional migrations using graphical buttons located on the main toolbar, directly within the JupyterHub environment.
This allows users to control the migration process as well as viewing HPC partitions' status without leaving the notebook interface, all while maintaining no user intervention.
The architecture of JuPyLive consists of four modules, the following subsections describe the implementation and functionality of each module in details.

\subsection{Frontend Integration}
To simplify the accessibility for users, the underlying functionalities are wrapped in two main logical blocks that can be accessed using Jupyter magic commands.
However, to further improve the accessibility, JuPyLive fully integrates into the Jupyter interface by adding intuitive visual buttons that invoke underlying functionalities such as “Migrate” and “HPC Status”.
This approach, minimizes the distractions and helps users to concentrate better on their own code, while keeping the migration functionality always conveniently visible from the toolbar.

Under the hood, the Migrate button triggers a callback function that executes the \verb|%MigrateTo| magic command, and similarly the HPC Status button triggers a callback to invoke the \verb|%HPCStatus| magic command.
A custom TypeScript\,\footnote{TypeScript is an open-source high-level programming language that is transpiled to standard JavaScript before execution so that it can be interpreted by the browser.} extension injects these functional buttons into the notebook toolbar at runtime. 
To bridge the frontend UI with the backend migration logic, the extension utilizes the JupyterLab \verb|ISessionContext| in order to send execution requests directly to the active IPython kernel.
This allows the buttons to capture kernel streams in real-time, enabling the UI to handle user inputs for partition selection and perform automatic browser redirection once the remote notebook server is ready after migration.
 
\subsection{Migration Engine}
The heart of JuPyLive is its migration mechanism that executes a sequence of coordinated steps to orchestrate checkpoint creation, resource allocation, data synchronization, and remote bootstrapping.
The sequence of steps depends on spatial condition of the notebook environment, i.e. being on the source (local workstation) or at destination (HPC cluster).


Based on the chosen migration target by user, the migration engine loads necessary components and imports the backend functions to handle check-pointing of current runtime context.
Upon execution of each cell in a notebook, all its events and variables are recorded~\cite{elasticNotebook}.
Then a checkpoint of the notebook is created.
This produces a portable snapshot containing variable contents as well as their relationships and dependencies.
Afterwards, the migration engine proceeds to establish a remote connection between source and destination for the environment synchronization.
Finally, the local project directory including the checkpoints is synchronized with the project directory on destination\,\footnote{Bulky directories such as node-modules and .env are excluded.}.
To ensure environment consistency, the migration engine executes a remote dependency alignment phase.
Before the initiation of the notebook server, the migration engine prepares a compatible environment in the user namespace on the remote host to synchronize references and to provide required dependencies.
This step is decoupled from the main bootstrapping process to allow for independent error handling and to ensure the readiness of the execution environment before that a user session begins.

In parallel with the above steps, the migration engine interacts with the batch job scheduler (Slurm) to obtain required resources on the remote machine (HPC node).
It first checks for existing allocations, and if none is available, it requests a new resource allocation.
The migration engine automatically includes relevant hardware-specific flags in the resource request query, to ensure that the necessary computing resources required for a task (e.g., GPU) are correctly allocated.

Once the HPC resources are granted, the migration engine initiates the bootstrapping process to prepare the environment.
This process injects a startup script into the IPython profile directory, ensuring that the remote session inherits the same configuration as the original instance.
It also handles the symlinking of the JupyterLab UI extension, such that all buttons from the local toolbar remain available and active on the remote interface, and registers predefined magic commands that enable the reverse migration upon users interaction.
As the final step, the bootstrap process kills any existing notebook instance on the assigned port and launches a new Jupyter server, while redirecting the user to the migrated notebook.
A reverse SSH tunnel binds the remote Jupyter port to a local port on users machine, allowing immediate reconnection of the user’s browser to the active session which is now fully migrated and runs on a remote HPC node.

While the remote Jupyter notebooks are in use, a background Pull Watcher thread remains running on the local workstation, this process enables instant bi-directional migration despite all network complexities.
This lightweight process periodically polls the HPC project directory for a specific signal file.
When a user initiates a return migration by pressing the Migrate button on the HPC side, the remote instance saves its state and creates this signal file.
Once the Pull Watcher detects the trigger (signal file), it initiates a reverse synchronization operation to pull the updated files and checkpoint back to the local folder, seamlessly restoring the execution state of local Jupyer notebook.
At this point, the user just needs to reload the browser tab and continue the development using the local resources.
The migration between various computing resources can be repeated as many times as required.


\subsection{HPC Telemetry and Decision Support}
\begin{figure}
    \centering
    \includegraphics[width=\textwidth]{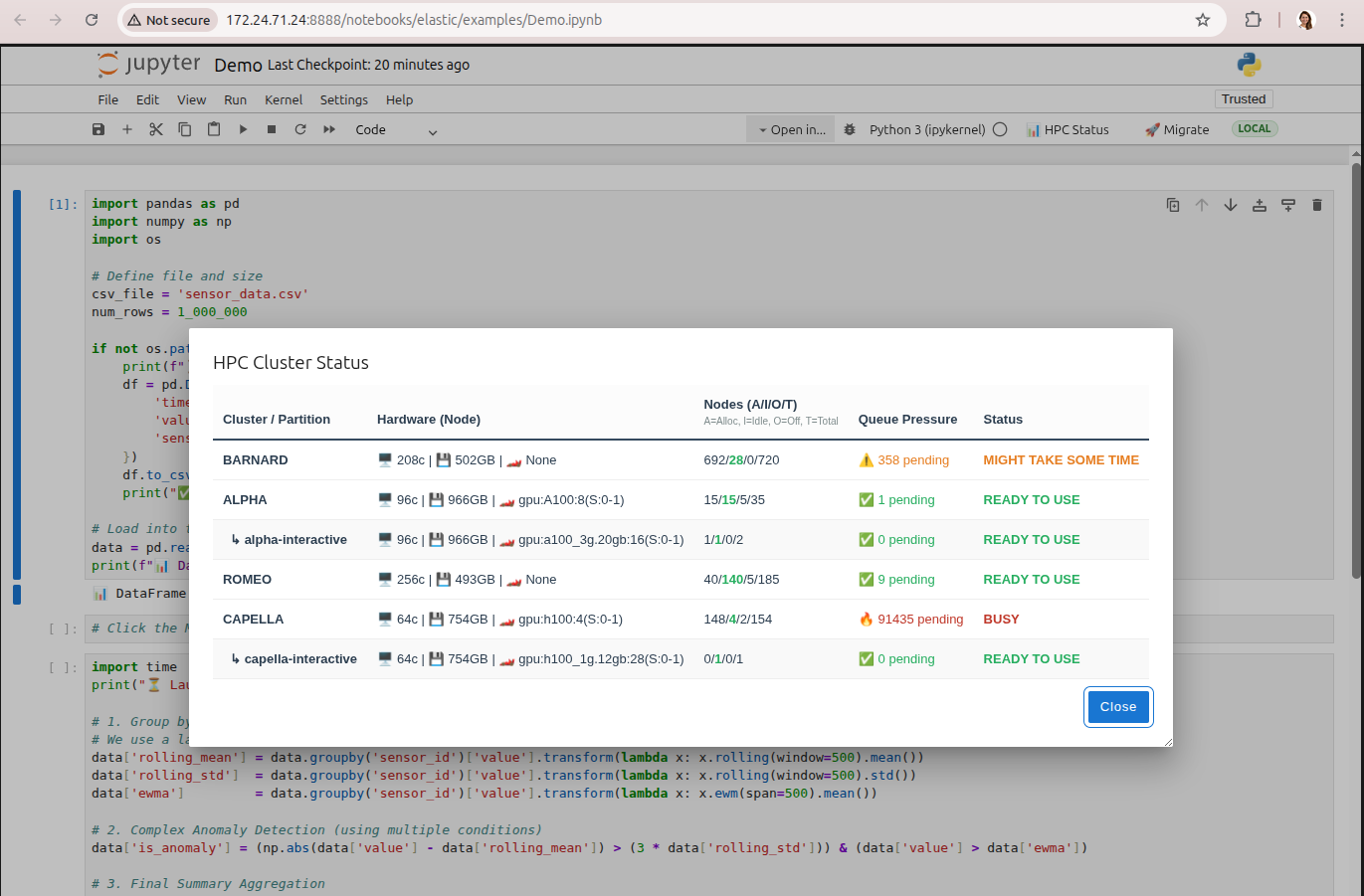}
    \caption{HPC status: queue depths and availability for each cluster}
    \label{fig:telemetry}
\end{figure}

Considering that HPC resources are limited and highly demanded, for certain processes, it might be more beneficial to continue the process on low performance resources instead of waiting in a queue to obtain HPC resources.
Therefore, JuPyLive not only enables seamless migration but also helps users decide where and when to perform a migration.
The “HPC Status” functionality, which is accessible from the navigation toolbar displays the real-time status of all connected HPC clusters using an intuitive graphical interface directly within the Jupyter notebook.
This function, under the hood,  uses a \verb|ThreadPoolExecutor| to fetch the relevant data concurrently for all clusters and aggregates the results accordingly.
The provided data shows for each partition, its resources, current availability, and the length of the waiting queue.
Furthermore, it computes a readiness assessment and presents a color‑coded recommendation for each cluster as shown in Figure~\ref{fig:telemetry}.
In this representation, BUSY resources are shown in red, those that MIGHT TAKE SOME TIME in orange, and resources that are READY TO USE are shown in green.

\subsection{Automatic Environment Preparation}
\label{sec:script}
Consistent with the central aim of this work, which is making high-performance computing accessible to everyone, an all-in-one installer is provided that performs a zero-configuration setup of JuPyLive.
This comprehensive installer handles all steps of environment preparation automatically, from system dependencies to secure remote access.
Thus, enabling user to immediately start coding in the familiar Jupyter notebook without any manual configuration steps, and gives them the freedom to seamlessly access the resources on any connected HPC cluster on demand.

The installer begins by creating a configuration file, while interactively  prompting for required information such as HPC cluster access credentials.
It then checks and installs all required system dependencies, and creates a clean Python virtual environment including all requirements and Jupyter-specific packages.
Next, it manages authentication by generating an RSA SSH key pair (if missing) and enabling a Key-based SSH connection to HPC clusters.

The installer then prepares the local Jupyter environment by creating the IPython startup directory and linking a custom extension.
It also cleans up any lingering SSH agent connections to prevent interference. 
Finally, the installer launches a headless Jupyter server bound to the host's IP address that automatically loads the main magic commands.

\subsection{Limitation and Assumptions}
\label{sec:assumptions} 
Current implementation considers few realistic assumptions.
The amount of requested resources on HPC clusters are currently predefined, which in future developments will be adjustable by users.

The existing implementation natively supports Slurm scheduler and the access points to the HPC clusters
are manually set in the configuration files.
Therefore, running JuPyLive on other platforms requires a one-time adaptation by respective HPC cluster administrators, the adapted configuration files then can be used by all users of that HPC cluster without further modifications.

Despite its apparent clarity, it worth mentioning that for successful migrations, the remote resource must be accessible from the local machine.

When a busy or semi-busy resource (red/orange) is selected, obtaining the required resources can take some time.
Once a migration eventually takes place, all changes made to the local notebook during this waiting time will be ignored.
Finally, migration of notebooks with large internal context may slow down the overall process.

\section{Conclusion and Future Direction}
\label{sec:conclusion} 
This work introduces JuPyLive, a practical migration mechanism between local workstations and HPC resources, that enables a seamless switch between computational resources on-demand.
This mechanism bridges the gap between interactive code development on local yet low performance resources and the high performance yet batch processing resources of the HPC clusters, automating the entire orchestration of states, files, and network routing.
While the current implementation is designed to demonstrate the feasibility of a “one-click” seamless migration among fully heterogeneous environments, it also serves as a foundation for sophisticated resource management strategies.
For future developments, as the first step the synchronization strategy will be optimized.
Additionally, utilizing a mountable shared storage between source and destination to minimize the data transfer is planned.
Currently hard coded resource requests will be replaced with dynamic allocation using resource prediction techniques to automatically match required resources.
Finally, historical migration patterns, queue conditions, and resource utilization records will be analyzed using machine learning techniques to proactively suggest optimal migration targets, further simplifying the decision process for users.

\begin{credits}
\subsubsection{\ackname}
This work was supported by the Federal Ministry of Education and Research of Germany and by the Sächsische Staatsministerium für Wissenschaft, Kultur und Tourismus in the programme Center of Excellence for AI-research “Center for Scalable Data Analytics and Artificial Intelligence Dresden/Leipzig”, project identification number: ScaDS.AI; and by the German Research Foundation (DFG) project NFDI4DataScience (no. 460234259). The funding bodies had no role in the design of the study, collection, analysis, and interpretation of data, or in writing the manuscript.
\end{credits}

\bibliographystyle{splncs04}
\bibliography{ref}
 
\end{document}